\documentclass[prb,aps,twocolumn]{revtex4-1}
\usepackage{bm}
\usepackage[colorlinks=true,linkcolor=blue,citecolor=blue]{hyperref}
\usepackage{times}
\usepackage{amsmath}
\usepackage{amssymb}
\usepackage{amsthm}
\usepackage{amsfonts}
\usepackage{enumerate}
\usepackage{latexsym}
\usepackage{ifpdf}

\usepackage{natbib}
\newcommand{\beq}{\begin{equation}}
\newcommand{\eeq}{\end{equation}}
\usepackage{graphicx}
\usepackage{makeidx}
\usepackage{color}
\usepackage{array}
\usepackage{multirow}

\begin{document}

\title{Anomalous electron transport in epitaxial NdNiO$_3$ films}

\author{Shashank Kumar Ojha}
\affiliation{Department of Physics, Indian Institute of Science, Bengaluru 560012, India}
\author{Sujay Ray}
\affiliation{Department of Physics, Indian Institute of Science, Bengaluru 560012, India}
\author {Tanmoy Das}
\affiliation{Department of Physics, Indian Institute of Science, Bengaluru 560012, India}
\author {S. Middey}
\email{smiddey@iisc.ac.in }
\affiliation{Department of Physics, Indian Institute of Science, Bengaluru 560012, India}
\author{Sagar Sarkar}
\affiliation  {S.N. Bose National Center for Basic Sciences, JD-Block, Sector III, Salt Lake, Kolkata 700098, India}
\author{Priya Mahadevan}
\affiliation  {S.N. Bose National Center for Basic Sciences, JD-Block, Sector III, Salt Lake, Kolkata 700098, India}
\author{Zhen Wang}
\affiliation{Department of Condensed Matter Physics and Materials Science, Brookhaven National Laboratory, Upton, NY, 11973, USA}
\author{Yimei Zhu}
\affiliation{Department of Condensed Matter Physics and Materials Science, Brookhaven National Laboratory, Upton, NY, 11973, USA}
\author{Xiaoran Liu}
\affiliation{Department of Physics and Astronomy, Rutgers University, Piscataway, New Jersey 08854, USA}
\author{M. Kareev}
\affiliation{Department of Physics and Astronomy, Rutgers University, Piscataway, New Jersey 08854, USA}
\author{ J. Chakhalian}
\affiliation{Department of Physics and Astronomy, Rutgers University, Piscataway, New Jersey 08854, USA}

\begin{abstract}
The origin of simultaneous  electronic, structural and magnetic transitions in  bulk rare-earth nickelates ($RE$NiO$_3$) remains puzzling  with multiple conflicting reports on the nature of these entangled phase transitions. Heterostructure engineering of these materials offers  unique opportunity to decouple metal-insulator transition (MIT) from the magnetic transition. However, the evolution of underlying electronic properties across these decoupled transitions remains largely unexplored. In order to address this, we have measured Hall effect  on  a series of epitaxial NdNiO$_3$ films, spanning a variety of  electronic and magnetic phases. We find that the MIT results in only partially gapped Fermi surface, whereas full insulating phase forms below the magnetic transition. In addition, we also find a systematic reduction of the Hall coefficient ($R_H$) in the metallic phase  of these films with epitaxial strain and also a surprising transition to negative value at large compressive strain. Partially gapped weakly insulating, paramagnetic phase is reminiscence of pseudogap behavior of high $T_c$ cuprates. The precursor metallic phase, which undergoes transition to insulating phase is a non-Fermi liquid with the temperature exponent ($n$) of resistivity of 1, whereas the exponent increases to 4/3 in the non-insulating samples. Such nickelate phase diagram with sign-reversal of $R_H$, pseudo-gap phase and non Fermi liquid behavior are intriguingly similar to high $T_c$ cuprates, giving important guideline to engineer unconventional superconductivity in oxide heterostructure.
 \end{abstract}

\maketitle

 \section{Introduction}

The question about the nature  of metal-insulator transition  (MIT) and spin ordering  in the negative charge transfer family of materials $RE$NiO$_3$ (\textit{RE}=Pr, Nd...Lu etc.)  has drawn a significant interest  in the pursuit of  understanding the ultimate connection among the underlying crystal structure, electronic, and magnetic orderings~\cite{Medarde:1997p1679,Catalan:2008p729,Middey:2016p305,Catalano:2018p046501,Mizokawa:2000p11263,Staub:2002p126402,Lee:2011p016405,Lee:2010p165119,Stewart:2011p176401,
Park:2012p156402,Johnston:2014p106404,Hepting:2014p227206,Upton:2015p036401,Subedi:2015p075128,Meyers:2016p27934,Green:2016p195127,Bisogni:2016p13017,Haule:2017p2045,Mandal:2017p06819,Lu:2018p031014,Middey:2018p156801,Mercy:2017p1677,Shamblin:2018p86}.   Independently,  there has been a number  of interesting theoretical proposals to  realize  high temperature superconductivity through epitaxial engineering~\cite{Chaloupka:2008p016404,Hansmann:2009p016401,Hansmann:2010p235123,Han:2011p206804,Han:2013p179904}, leading  to the remarkable progress in synthesis and characterization of ultra-thin film  and heterostructures of rare-earth nickelates  (for recent progress see Refs.~\onlinecite{Middey:2016p305,Catalano:2018p046501}  for review and the references therein). Since  the degeneracy lifting between two $e_g$ orbitals of Ni$^{3+}$ ions  might lead to  cuprate-like one band Fermi surface, orbital engineering of $RE$NiO$_3$  have been also attempted in various heterostructure forms~\cite{Middey:2016p305,Benckiser:2011p189,Freeland:2011p57004,Chakhalian:2011p116805,Wu:2013p125124,Disa:2015p062303,Middey:2016p056801}. However, to-date maximum achieved orbital polarization of  nickelate heterostructures ($\sim$25\%) is still significantly smaller than  that which is required to architecture the materials analog of high $T_c$ cuprates.  Therefore, it is of  a paramount interest to find out if and how epitaxy can be utilized  to finally achieve cuprate like electronic structure in $RE$NiO$_3$ based heterostructures.

Hall effect is  an important measurement which  provides crucial information about the Fermi surface topology,  the carrier concentration, the anisotropy of scattering rate, and the chiral spin textures of quantum materials~\cite{Ong:1997p977,Ando:2004p197001,Ono:2007p024515,Nagaosa:2013p899}.    In connection to nickelates, earlier studies  on bulk NdNiO$_3$ (NNO)\ and PrNiO$_3$ (PNO)\ powder samples demonstrated  the phenomenon of  sign change  of Hall coefficient ($R_H$) across the metal-insulator transition~\cite{Cheong:1994p1087}. Interestingly, such evolution of $R_H$ has been also observed in several important `bad' metals including cuprates, vanadates, and ruthenates; this phenomenon  was attributed to  multiple   factors including  structural transition, spin density wave (SDW) transition, charge density wave (CDW) transition~\cite{LeBoeuf:2007p533, Kim:2006p266401, Galvin:2001p161102, Xing:2018p041113}.  Therefore, the simultaneous occurrence of MIT, structural transition, charge  (CO) and magnetic ordering in bulk NNO and PNO~\cite{Staub:2002p126402,Lorenzo:2005p045128,Scagnoli:2005p155111,Scagnoli:2006p100409} inhibits the straightforward determination of the primary  factor responsible for such drastic change in $R_H$.  On the other hand, these simultaneous transitions can be selectively decoupled or  even suppressed  by  epitaxial strain in ultra-thin film geometry~\cite{Liu:2013p2714,Mikheev:2015p1500797,Hepting:2014p227206,Hauser:2015p092104,Upton:2015p036401,Meyers:2016p27934}. Naturally,  in such situation one can expect that the Hall effect measurement will   provide crucial information about the  electronic and magnetic transitions. Further, it is  interesting  to note that epitaxial strain  dramatically affects the electronic properties in the metallic phase of nickelate films as highlighted by the change in the characteristic power-exponents  of resistivity, associated with the non-Fermi liquid behavior~\cite{Liu:2013p2714,Mikheev:2015p1500797,Middey:2018ENOLNOStrain}.   This observation suggests that Hall effect measurements in the metallic phase  can probe the relative change in Fermi surface topology brought   about by   epitaxial strain~\cite{Lee:2018p227}.

   In this paper, we report on the detailed Hall effect measurements across  several  electronic and magnetic phases of NNO, which have been realized in a series of epitaxially stabilized  high-quality ultra-thin films  on several single crystalline substrates. These measurements  have revealed a direct link between the sign change of $R_H$ and  the onset of $E^\prime$  antiferromagnetic ($E^\prime$-AFM~\cite{Scagnoli:2006p100409}) transition implying  a SDW (spin density wave) origin of the puzzling $E'$-AFM phase. Such magnetic transition driven by the Fermi surface reconstruction  has been  predicted by earlier theoretical works~\cite{Lee:2011p016405,Lee:2010p165119}.  Moreover,  metallic phase of the NNO film under large tensile strain exhibits unexpected cuprate-like linearly $T$-dependent resistivity and $T^2$ dependence of cotangent of Hall angle. The magnitude of  $R_H$  in the metallic phase  shows a systematic decrease with the  underlying  strain  and remains positive down to lowest temperatures under moderate compressive strain.  Upon application of large compressive strain,  surprisingly   negative $R_H$ ($T>$ 60 K) emerges in the metallic phase,  emphasizing a drastic change in Fermi surface topology. The decrease (increase) in hole (electron) concentration without any chemical doping illustrates a strain mediated self-doping scenario, which is further verified  by  density functional theory (DFT). Our further analysis using DFT+MRDF (momentum resolved density fluctuation) method found that the suppression  of $E^\prime$-AFM ordering by epitaxial strain can be accounted by the  suppression of  Fermi surface nesting.

   \begin{figure}[b!]
\vspace{-0pt}
\includegraphics[width=0.48\textwidth] {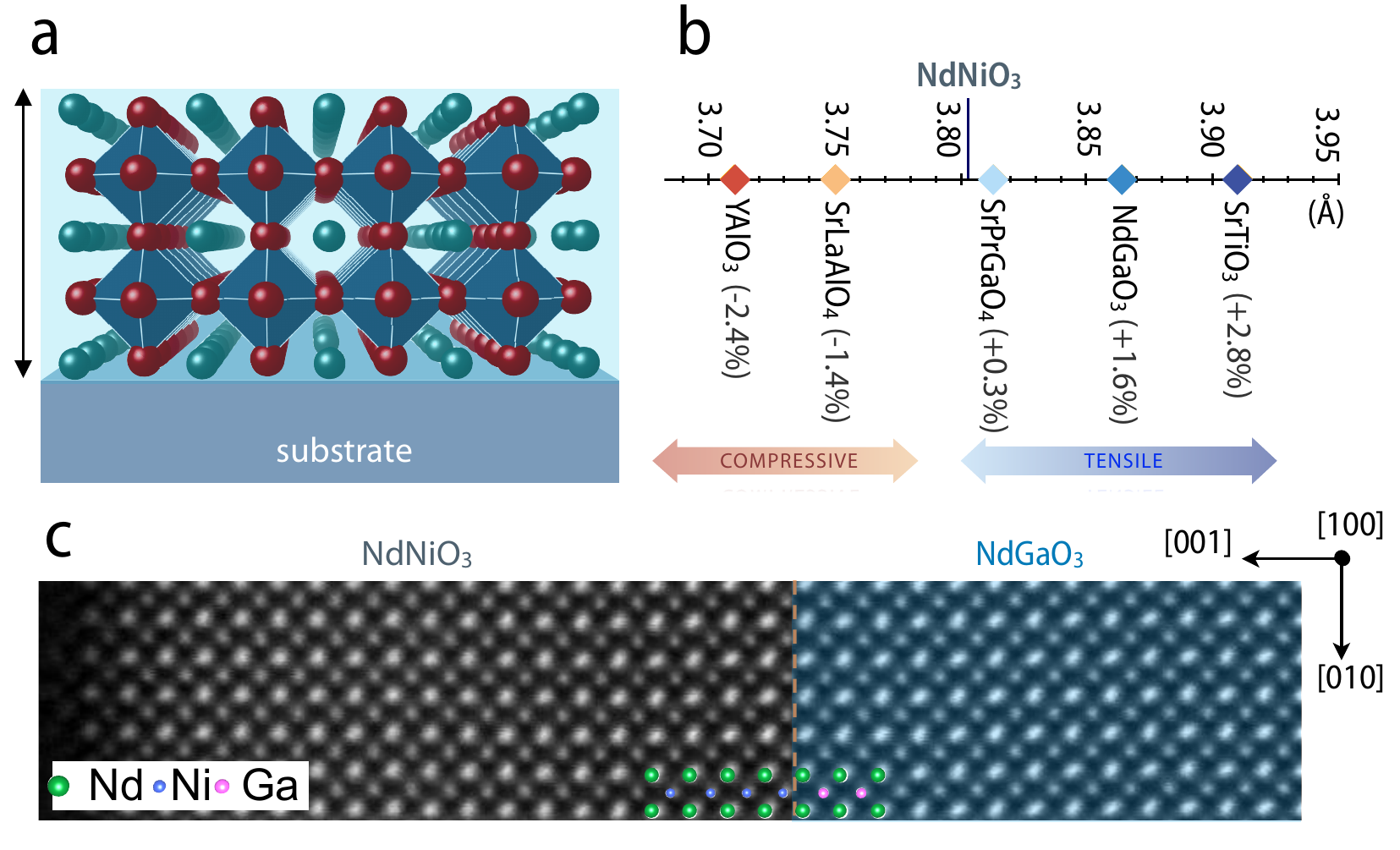}
\caption{\label{Fig1} (a) NdNiO$_3$ thin film on single crystalline substrate. (b) Pseudocubic in-plane lattice constant for the substrates used in this work and the corresponding epitaxial strain for NNO. (c) HAADF-STEM image (with false color) of a 20 u.c. NdNiO$_3$ film on NdGaO$_3$ substrate. }
\end{figure}

\section{Methods}
Ultra-thin films of NNO   (15 unit cell=u.c.) of  high structural  and chemical quality have been grown on several single crystalline substrates [see Fig. \ \ref{Fig1}(a)]:  SrTiO$_3$ (STO), NdGaO$_3$ (NGO), SrPrGaO$_4$ (SPGO), SrLaAlO$_4$ (SLAO) and YAlO$_3$ (YAO). The details of the growth procedure can be found in Ref.~\onlinecite{Liu:2013p2714}.  The in-plane pseudocubic lattice constants of all used  substrates and  the corresponding strain ($\epsilon$) values  are summarized in Fig.\ \ref{Fig1}(b).  The  layer-by-layer growth has been monitored by in-situ RHEED (see supplemental~\cite{supplemental})  and the desired epitaxial growth along  a pseudo-cubic [0 0 1] direction has been confirmed by X-ray diffraction~\cite{supplemental}. A sample for cross-section STEM (scanning transmission electron microscopy)  measurement has been prepared by a focused ion beam with Ga$^{+}$ ions followed by Ar$^{+}$ ions nano-milling. High-angle annular dark-field (HAADF) imaging has been achieved with a JEOL ARM200 microscope equipped with two aberration correctors.  Figure~\ref{Fig1}(c) shows a characteristic HAADF-STEM image taken across the NNO/NGO interface  along [1 0 0] direction; The atomic column intensity in the HAADF-STEM imaging varies with the atomic number as $\propto Z^{1.7}$, i.e.   heavier  atoms indicate brighter  columns. As marked by the dashed line, the NGO substrate terminates with the NdO layer with practically atomically sharp interface which further testifies for the excellent registry  between the sample and substrate. $dc$ resistivity and Hall  effect measurements have been carried out in  four-probe  van der Pauw geometry using a Quantum Design PPMS (Physical Property Measurement System). For evaluating $R_H$ magnetic field ($H$) has been swept  between $\pm$5 T  at different $T$. In the absence of anomalous Hall contribution, while the intrinsic Hall resistance ($R_{xy}$) should be zero at  $H$ = 0 and asymmetric with magnetic field sweep, the finite  width of contacts adds an additional symmetric  part in $R_{xy}$ about $H$ = 0 with   a vertical offset~\cite{Jones:1993p8986,Hauser:2013p182105}. These parasitic contributions were corrected to extract intrinsic $R_{xy}$ and   three typical sets of such corrected  $R_{xy}$ curves as a function  of  $H$  for different $T$ are  presented in the Supplemental ~\cite{supplemental}. As clearly seen, after correction  $R_{xy}$ remains linear within the ranges of magnetic field ($H$)  used in this work;  Hall coefficient   has been evaluated as $R_H$ = $t (dR_{xy}/dH)$ where $t$ is the film thickness.

To  gain further  insight of  the change in $R_H$ in the metallic phase of the samples as a function of epitaxial strain, we have performed density functional theory (DFT) calculation using the Vienna {\it ab-initio} simulation package~\cite{Kresse:1996p15,Kresse:1996p11169} within the GGA+$U$ of PBE parametrization~\cite{Perdew:1996p3865}. Projected augmented-wave (PAW)~\cite{Blochl:1994p17953,Kresse:1999p1758}~pseudo-potentials were used to describe core electrons. We use $U=3.5$~eV, which is larger than the values of $U$ used in the self-energy calculation. This is expected since the bare $U$ used in the self-energy calculation is further multiplied by various components of the susceptibility to provide the effective many-body potential in this calculation as detailed below. The electronic wave-function is expanded using plane waves up to a cutoff energy of 500 eV. Brillouin zone (BZ) sampling is done by using a (12$\times$12$\times$6) Monkhorst-Pack $k$-grid. Similar to Ref.~\onlinecite{Liu:2013p2714},   NdNiO$_3$ crystal structures have been constrained to $P4/mmm$ space group in our calculation and $A$ type antiferromagnetic spin ordering on Ni sublattice has been imposed, instead of complex  $E'$-AFM spin configuration.   Octahedral tilts/rotations and breathing mode distortions have been also omitted in our calculations and the effect of substrates have been folded in to  the experimental lattice constants. Despite these constraints, such DFT  approach was shown  to successfully reproduce the experimentally observed band structure and Fermi surface topology of NNO thin films on different substrates~\cite{Dhaka:2015p035127}. The effects of electronic correlations on the  band structure have been investigated  by DFT and DFT+self energy corrections, obtained by  MRDF theory (details of this calculation are in Supplemental Information)~\cite{Das:2012p017001,Das:2015p094510,Das:2014p151,Dhaka:2015p035127}.

\section{Results and Discussions}

\begin{figure*}
\vspace{-0pt}
\includegraphics[width=0.95\textwidth] {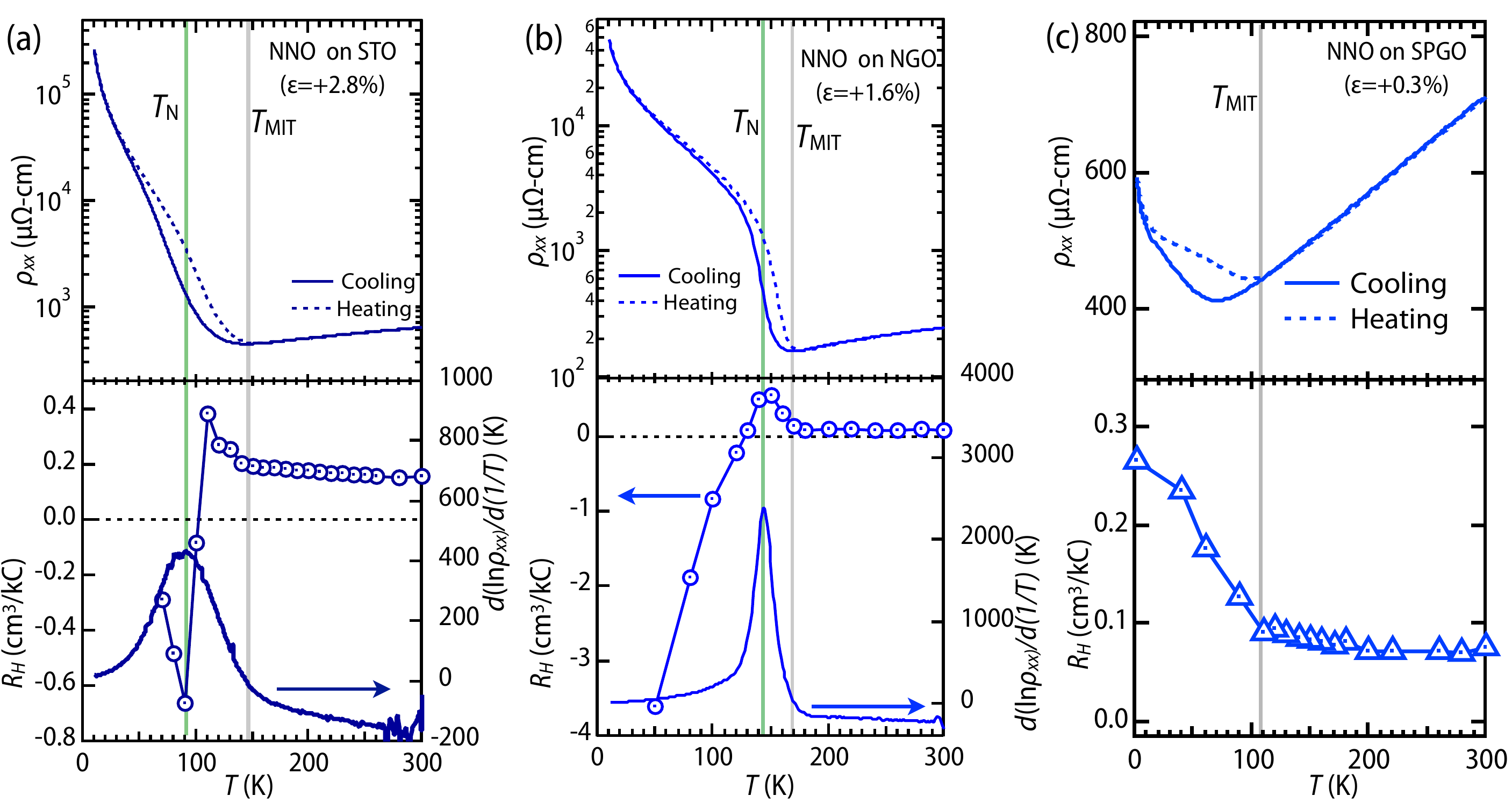}
\caption{\label{Fig2}  Temperature dependence of \textit{dc} resistivity  of 15 u.c. NNO film under tensile strain  in upper panel of (a)-(c). The variation of  $R_H$ are shown in corresponding  lower panel. Right axis of the lower panel of (a) and (b) corresponds to $d(\mathrm {ln}\rho_{xx})/d(1/T)$ $vs$.  $T$ plot. }
\end{figure*}

\subsection{Hall coefficient across electronic and magnetic transitions}  The upper panels of Fig.\ \ref{Fig2}(a)-(c)  show the temperature dependent resistivity of NNO films under tensile strain. As reported earlier~\cite{Liu:2013p2714,Mikheev:2015p1500797}, NNO  films  under tensile strain  show  the first order MIT.  Metal-insulator transition temperature $ T_{\mathrm{MIT}}$ [defined as the temperature where $d\rho_{xx}/dT$=0] and magnetic transition temperature $T_N$ [evaluated from $d(\mathrm {ln}\rho_{xx})/d(1/T)$ $vs$.  $T$ plot (right axis of lower panel of Fig. \ \ref{Fig2}(a)-(b))~\cite{Zhou:2005p226602} ] of these films are listed in Table-I. Note, lower value of  $ T_{\mathrm{MIT}}$ compared to the bulk NNO $T_{\mathrm{MIT}} = T_N \sim$ 200 K   were linked to the effect of epitaxial strain (bandwidth control) and reduced dimensionality (quantum confinement) ~\cite{Liu:2013p2714}.  Moreover, previous reports using resonant X-ray  scattering (RXS) on such NNO films have confirmed the absence of both bulk-like charge ordering and lattice symmetry change across the MIT~\cite{Upton:2015p036401,Meyers:2016p27934}. Separation between $T_{MIT}$ and $T_N$  of these samples offers  a unique temperature window to examine the evolution  of $R_H$ across all three phases: paramagnetic metal (PM), paramagnetic insulator (PI) and antiferromagnetic  insulator (AFI) without influence of charge disproportionation (CD)\ and structural transition. While a bulk-like MIT and magnetic transition is expected in epitaxial NNO film  under very   small tensile strain of +0.3\%,  surprisingly   the film shows  only weakly insulating, paramagnetic behavior at low temperature~\cite{Liu:2013p2714} [see upper panel in Fig. \ \ref{Fig2}(c)].

\begin{table}
\caption {Metal-insulator transition temperature ($T_\mathrm{MIT}$ ) and antiferromagnetic ordering temperature ($T_N$), evaluated from $\rho_{xx}$ vs. $T$ data. $c$ and $h$ correspond to cooling and heating cycle, respectively.}
%\begin{tabular}{|c|c|c|c|c|}

\begin{tabular}{lcccc}
\hline
\hline
 Film     & $T^c_\mathrm{MIT}$     &  $T^h_\mathrm{MIT}$     & $T^c_N$&  $T^h_N$ \\
                    \hline
NdNiO$_3$ on SrTiO$_3$ & 145 K & 155 K &  90$\pm$5 K &  110$\pm$5 K       \\
 NdNiO$_3$ on NdGaO$_3$ & 170 K & 175 K &  140 K  &  155 K  \\  \hline
\end{tabular}
\end{table}

In the following, we discuss overall response of $R_H$ across the electronic and magnetic transitions of these films. The temperature dependence of  $\rho_{xx}$ and $R_H$  in the metallic phase  will be presented latter in the text.  As immediately seen in Fig.\ \ref{Fig2}(a)-(c) at  room temperature  $R_H$ is hole-like.   Though the electronic structure of the system is expected to be drastically different across the metal-insulator transition, $R_H$  exhibits only a slight increase  across $T_\mathrm{MIT}$ for the NNO film on NGO and STO. Most interestingly, $R_H$ switches to n-type  around 100 K and 120 K for STO and NGO cases respectively, which are remarkably close to the respective $T_N$ for $E'$-AFM ordering.  Similar sign change of $R_H$ across $T_N$ has been also observed in 2EuNiO$_3$/1LaNiO$_3$ superlattice, which has monoclinic symmetry in both metallic and insulating phase and does not exhibit any charge ordering transition~\cite{Middey:2018p156801}.  $R_H$  maintains  hole-like behavior even in the weakly insulating state for the film grown on SPGO substrate, where earlier RXS experiments~\cite{Liu:2013p2714}  had clearly ruled out the appearance of $E'$-type magnetic ordering. All of these observations clearly point at  some large changes in the Fermi surface topology to be responsible for the appearance for  $E'$-type antiferromagnetic ordering in these materials.

  The mechanisms  of  the MIT and $E^\prime$ type AFM ordering of $RE$NiO$_3$ is  still heavily debated~\cite{Mizokawa:2000p11263,Staub:2002p126402,Lee:2011p016405,Lee:2010p165119,Stewart:2011p176401,
Park:2012p156402,Johnston:2014p106404,Hepting:2014p227206,Upton:2015p036401,Subedi:2015p075128,Meyers:2016p27934,Green:2016p195127,Bisogni:2016p13017,Haule:2017p2045,Lu:2018p031014,Middey:2018p156801,Mercy:2017p1677,Shamblin:2018p86}.    Most  recently, the   MIT mechanism has been attributed to the $d^8\underline{L}$+ $d^8\underline{L}\rightarrow d^8$+$d^8\underline{L}^2$ bond disproportionation (BD)  transition (here $\underline{L}$ denotes a ligand hole in O $p$ bands)~\cite{Mizokawa:2000p11263,Park:2012p156402,Johnston:2014p106404,Subedi:2015p075128,Bisogni:2016p13017,Middey:2018p156801,Ruppen:2015p155145}. Apart from the  BD induced transition scenario, the importance of Mott physics in realization of the insulating phase has been  found in optical  conductivity measurements~\cite{Stewart:2011p176401,Ruppen:2015p155145}.  Another unexpected result was revealed in the recent valence band photoemission measurement  which showed the presence of residual intensity at $E_F$ even in very low T thus signalling that  some parts of the Fermi surface still survive deep into the insulating phase~\cite{Schwier:2012p195147}. Such notion of a partially gapped Fermi surface, akin to the pseudogap phase of high $T_c$ cuprate, ruthenates, pnictides etc.~\cite{Damascelli:2003p473,Lee:2007p097403,Shimojima:2014p045101} was also inferred for NNO from IR spectroscopy~\cite{Stewart:2011p176401} and tunneling spectroscopy measurement~\cite{Allen:2015p062503}.

Since $R_H$ is magnetic field independent within the range of magnetic field used in this present study it  can be expressed as $R_H=(n_p\mu_p^2 - n_e\mu_e^2)/e(n_p\mu_p+n_e\mu_e)^2$, assuming one electron band and one hole band are contributing to the electrical transport, where $n_p$ ($n_e$) is the hole (electron) density and $\mu_p$ ($\mu_e$) is the mobility of the hole (electron)~\cite{Ha:2013p125150}.  The positive sign of $R_H$ in the paramagnetic metallic phases~\cite{Hauser:2013p182105,Son:2010p062114,Hauser:2015p092104} originates from the larger volume of the hole pockets  ($v_p$) compared to the electron pockets ($v_e$) as observed in    ARPES experiments~\cite{Eguchi:2009p115122,Dhaka:2015p035127}. Small increase of $R_H$  upon entering into the insulating phase suggests  that the BD transition opens only a partial gap in  Fermi surface.  ARPES measurement~\cite{Eguchi:2009p115122} found a strong  instability at  {\bf q} = (1/4, 1/4, 1/4)$_\mathrm{p.c.}$, which also coincides with the magnetic wave-vector for $E^\prime$  ordering. Thus  the nesting of   dominant hole pockets  in BD phase opens gap in the hole Fermi surface and results in the emergence  of  $E'$-AFM phase~\cite{Hepting:2014p227206,Hauser:2013p182105,Hauser:2015p092104,Lee:2011p016405,Lee:2010p165119,Yoo:2015p8746} with the remaining electron pockets contributing to the transport to  result  the observed switching of $R_H$ across $T_N$.  Once  BD phase and $E'$-AFM ordering set in , they act synergetically  to grow together with the decrease of thermal fluctuations~\cite{Ruppen:2017p045120}.  Complimentary experiments, like ARPES measurements at different $T$ are required to fully understand  the evolution of $R_H$ with $T$ in AFI phase. On  lowering of tensile strain, the suppression of the Fermi surface superstructure at (1/4, 1/4, 1/4)$_\mathrm{p.c.}$ is  found    to  be responsible for  the absence of $E^\prime$-AFM phase in NNO film on SPGO substrate where $R_H$ remains positive in such paramagnetic BD phase (results are shown in the Supplemental~\cite{supplemental}).

\begin{figure}
\vspace{-0pt}
\includegraphics[width=0.48\textwidth] {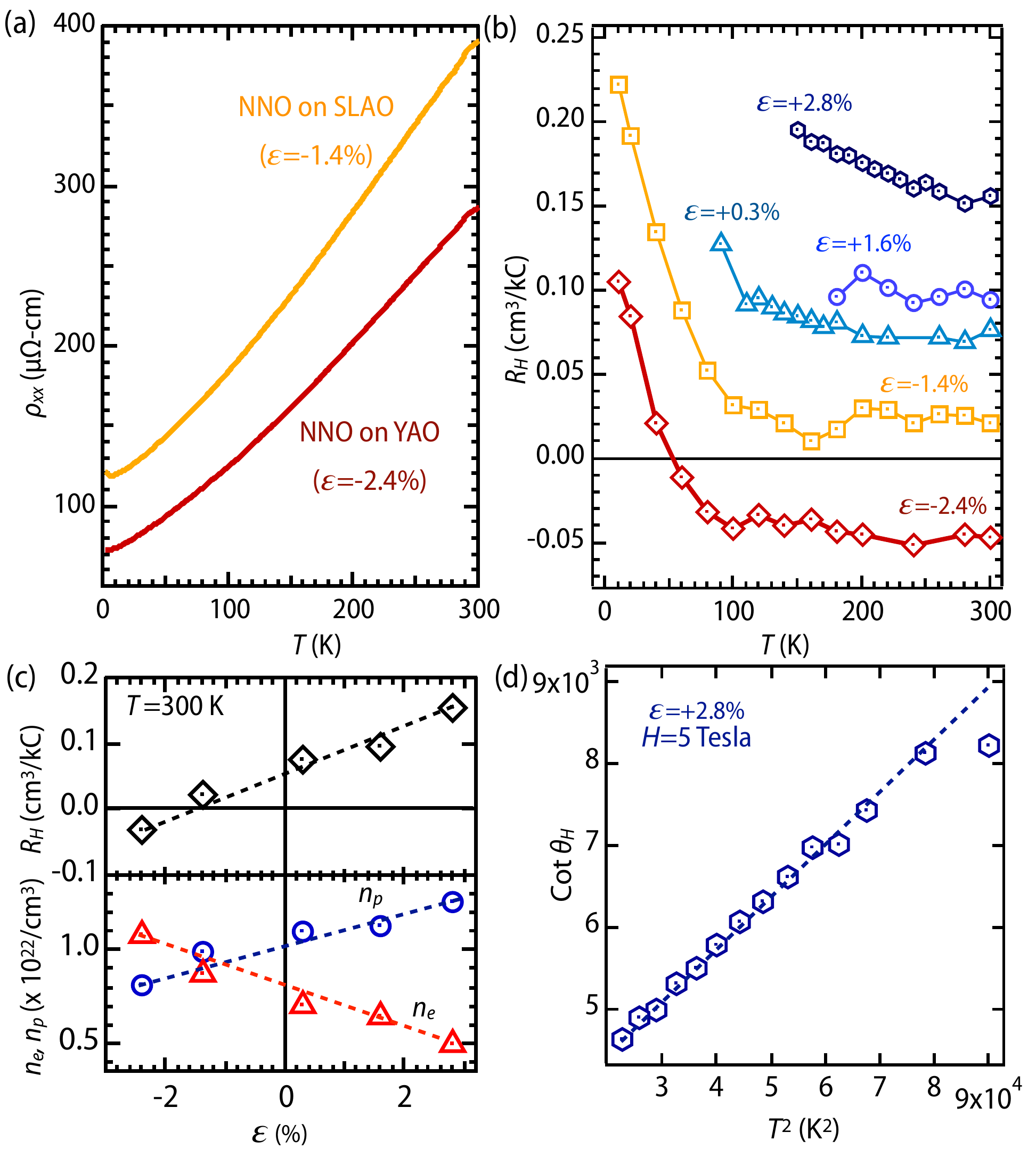}
\caption{\label{Fig3} (a) Temperature dependence of  \textit{dc} resistivity for 15 u.c. NNO film under compressive strain. (b) Temperature dependence of $R_H$ in the paramagnetic metallic phase of the NNO films. (c) Relation  of   $R_H$ (upper panel), carrier density (lower panel) with strain at 300 K. (d)  $\cot \theta_H$ for the NNO film on STO substrate as a function of $T^2$. The dotted line represents  $T^2$ dependence. }
\end{figure}

\subsection{$\rho_{xx}$ and $R_H$  in metallic phase}
Next  we explore  magneto-transport behavior in the metallic phase. Contrary to  the expected   $T^2$ dependence of  resistivity at low temperature for a Fermi liquid, $\rho_{xx}$ in  the metallic phases of NNO under  tensile strain    show  \textit{linear} $T$ dependent behavior (see Supplemental~\cite{supplemental}). As reported earlier~\cite{Liu:2013p2714}, the electronic and magnetic transitions can be entirely suppressed by the application of compressive strain [see Fig.\ \ref{Fig3}(a)].  Interestingly, under compressive strain  $\rho_{xx}$  shows    $T^{4/3}$ dependence  over a broad range of temperatures and then switches to linear $T$ behavior~\cite{supplemental}. Such non Fermi liquid behavior has been  observed  in the normal phase of several unconventional superconductors including cuprates, organic superconductor, pnictides, heavy-fermions etc.~\cite{Gegenwart:2008p186,Doiron:2009p214531,Taillefer:2010p51,Stewart:2011p1589} and its intrinsic  origin  is still unknown~\cite{Keimer:2015p179,Banerjee:2018}.

The overall behavior of $R_H$ in the metallic phase is  summarized in Fig. \ref{Fig3}(b).  As seen, the NNO film grown on SLAO substrate (-1.2\%) exhibits a $p$-type metallic behavior over the entire temperature range. With the further increase of compressive strain (NNO on YAO), $R_H$ surprisingly becomes negative even at 300 K. A systematic  relation between $R_H |_{T\mathrm{ = 300 K}}$ and $\epsilon$ can be further inferred  from the upper panel of   Fig. \ref{Fig3}(c).  Such strain mediated change in carrier density shown in  Fig. \ref{Fig3}(c)  is  a hallmark of the self-doping effect~\cite{supplemental}.

 \begin{figure*}[t!]
\vspace{-0pt}
\includegraphics[width=.75\textwidth] {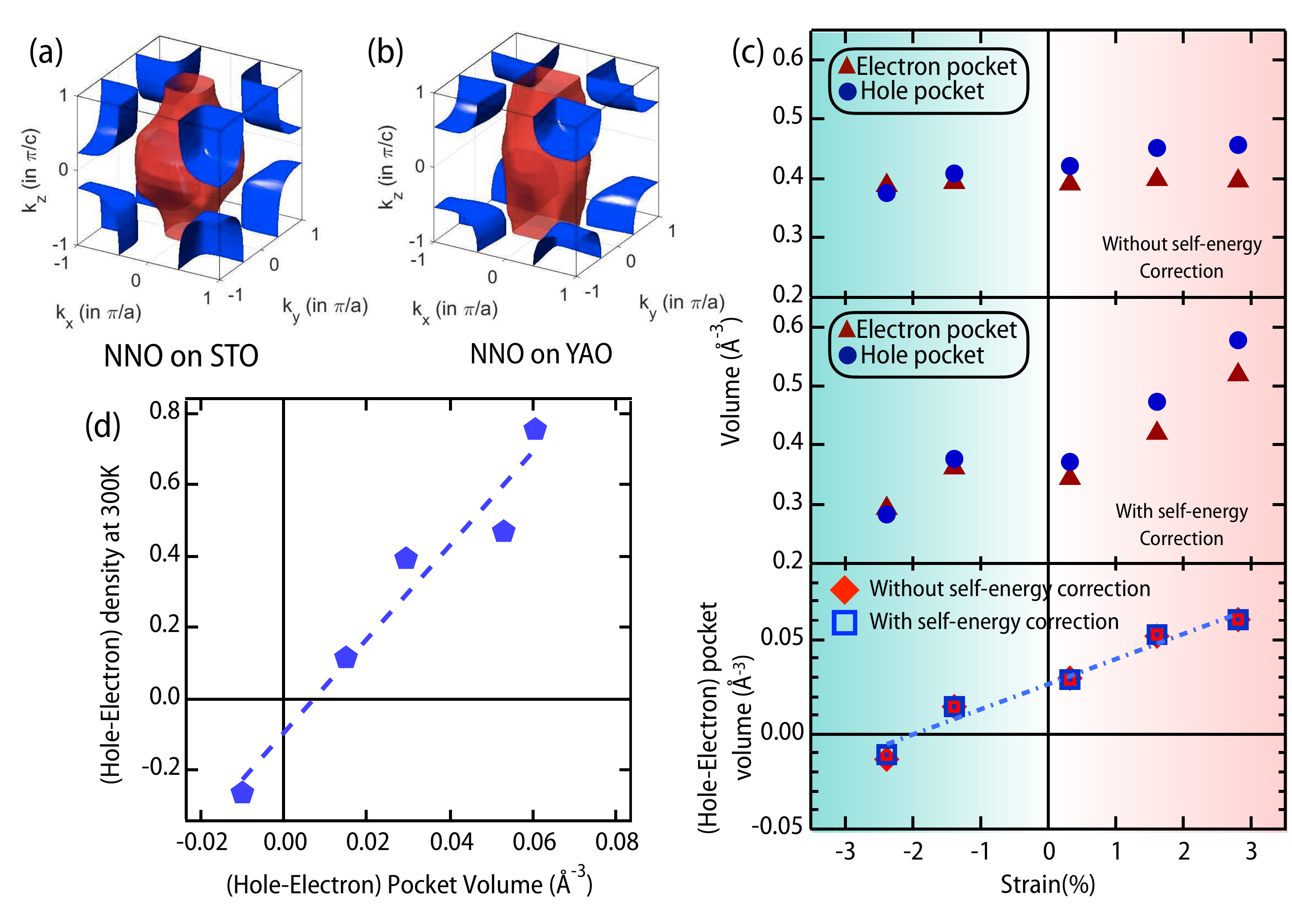}
\caption{\label{} (a)-(b) Computed FS topology are shown for the two representative samples of  NNO/STO and NNO/YAO. $a$ in $x$ and $y$ axes of these graphs correspond to in-plane pseudo cubic lattice constant of the substrate. For $z$ axis, $c\sim$2$c_\mathrm{p.c.}$, where $c_\mathrm{p.c.}$  is out-of plane lattice constant of NNO film. Details of Brillouin zone have been discussed in Supplemental information~\cite{supplemental}. (c) FS volume for electron (triangle) and hole (circles) pockets are compared for all samples without (top panel) and with (middle) self-energy corrections. Bottom panel compares the difference between the hole and electron pocket volume for calculations without (filled diamond) and with (open square) self-energy effects. (d) Comparison of the difference between the two FS volume with the experimental value of the effective carrier density obtained from Hall effect.}
\end{figure*}

For a normal metal, $R_H$ generally becomes temperature independent for $T>$  0.2-0.4 $\Theta_D$ ($\Theta_D$ is Debye temperature)~\cite{Chien:1991p2088}.   Since for $RE$NiO$_3$~$\Theta_D\sim$ 420 K \cite{Rajeev:1991p591}, strong $T$ dependent $R_H$ in the metallic phase of NNO film on STO requires some special consideration. For this purpose we recap that in hole doped cuprates the strong $T$ dependence of $R_H$  is commonly discussed in terms  of the Hall angle  $\cot \theta_H$=$\rho_{xx}$/($HR_H$).  The $T^2$ dependence of the $\cot \theta_H$, observed in hole-doped cuprates  around optimal doping has been interpreted  as a signature of spin-charge separation or result of anisotropic scattering rates~\cite{Anderson:1991p2092,Chien:1991p2088,Harris:1994p3246,Li:2016p197001,Stemmer:2018p062502}. However, $R_H$ becomes $T$-independent in lightly hole doped cuprates and consequently $\cot \theta_H$  follows the $T$ dependence of  $\rho_{ab}$~\cite{Ando:2004p197001}. Based on these results, we can speculate that  $T^2$ dependence of $\cot \theta_H$  [Fig.\ \ref{Fig3}(c)] and linear $T$ dependence of $\rho_{xx}$ for the NNO film on STO implies a strong connection with the `strange metal' phase of high $T_c$ cuprates.  As shown in Fig. \ref{Fig3}(b), with the decrease of hole concentration,  $R_H$ becomes almost \textit{T} independent in the metallic phase of NNO film on NGO and at  high temperatures for other compressive substrates.  Such sharp increase in $R_H$ below 100 K in compressively strained NNO  can be attributed to the emergence of the pseudogap phase~\cite{Allen:2015p062503,Uchida:2012p165126}, which is  observed in many cuprates and yet a poorly understood phenomena~\cite{Ando:2004p197001,Harris:1994p3246,Onose:2001p217001,Badoux:2016p210,Boyack:2019p134504}. In addition, the sign change of $R_H$ in pseudogap phase suggests that there is coexistence of electron and hole Fermi surface pockets with different mobility, as also seen in electron and hole doped cuprates~\cite{Dagan:2007p024506,Dagan:2004p167001,LeBoeuf:2007p533,Das:2008p2963}. Our result further implies that the hole pocket has higher mobility compared to electron-pocket, which is consistent with   prior ARPES data~\cite{Eguchi:2009p115122}.

\subsection{ DFT calculation}

In order to gain insight into the strain induced modulation of $R_H$ in the metallic phase shown in Fig. 3(c), we have calculated Fermi surface (FS) of NNO  for different values of $\epsilon$, listed in Fig. 1(b).  In all cases,  we obtain a characteristic FS topology which consists of the  electron-pocket at the center of the BZ and hole pockets at the zone corners. The computed FSs shown in Figs. 4(a) and 4(b) for large tensile (STO) and large compressive (YAO) strain exhibit the characteristic reduction of  the 3-dimensionality electron pocket (electron pocket is almost cylindrical for NNO on YAO) as a function of strain. In Figure 4(c), we summarize the evolution of FS volumes for the two pockets. As seen,  despite the gradual loss of three-dimensionality in Fermi surface between NNO under +2.8\% vs.  NNO under  -2.4\%, the electron pocket volume remains essentially unchanged. On the other hand, the hole pocket volume is larger than electron pocket volume for all cases except for NNO on YAO. This  is the key mechanism for the sign change in the Hall coefficient at  high temperatures where paramagnetic metallic state appears in all samples.

Since Mott physics is clearly important in driving antiferromagnetic and  insulating phases of nickelates~\cite{Stewart:2011p176401,Mercy:2017p1677}, we have further tested  the above  DFT results with the inclusion of  the self-energy correction due to electron-electron correlations. We compute the complex self-energy effects due to density-density fluctuations within the DFT+ MRDF theory~\cite{Das:2012p017001,Das:2015p094510,Das:2014p151,Dhaka:2015p035127}. The MRDF method directly incorporates the materials-specific DFT band structure and solve the Hubbard model (with intra and inter band Hubbard interactions) for the electron-electron correlation part; The electron correlations which arise due to  spin and charge density fluctuations are computed within the random-phase approximation. The feedback effects of the correlation to the electronic spectrum is captured within the  fluctuation-exchange (FLEX) method and quantified by band ($\nu$), momentum (${\bf k}$) and frequency ($\omega$) dependent complex self-energy $\Sigma_{\nu}({\bf k},\omega)$ corrections. We invoke self-consistency in such a way that the spin and charge correlation functions and the electronic Green's function include the self-energy corrections until the convergence in the self-energy value is reached~\cite{supplemental}.

The interacting Fermi surface is defined by the poles of the interacting Green's function at $\omega =0$ which is obtained  from the self-consistent solution of
$\varepsilon^{\nu}_{\bf k_{F}}+\Sigma^{\prime}_{\nu}({\bf k_{F}},\omega=0)=0$ where $\varepsilon_{\bf k}^{\nu}$ denotes the $\nu^{th}$ DFT band at momentum ${\bf k}$ and $\Sigma^{\prime}_{\nu}$ represent the real parts of the self-energy. We note that since self-energy is momentum dependent, the above solution can lead to a band-dependent change in shape of the Fermi surface. In what follows the ${\bf k}-$dependent self-energy causes a non-rigid-band shift of the Fermi surface, yet the total Fermi surface area remaining unchanged. This FS volume calculated with the self-energy dressed bands is shown in Fig. 4(c) middle panel). As seen, we obtain a non-monotonic behavior of the FS volume across different samples. However, in all cases, we find a one-to-one correspondence to the non-interacting FS volume, in that the electron pocket area is lower than that of the hole -pocket one, except for NNO on YAO. Thus, strain induced $n$-type metallic phase of NNO film results from the  change in relative FS volume by compressive epitaxy.

To further testify that the  estimated effective FS volume ($\delta v_\mathrm{FS}$)  truly correspond to the experimentally observed switching in $R_H$ as a function of $\epsilon$,  we plot $\delta v_\mathrm{FS}$ as a function of the difference in carrier concentration, $\delta n$=$n_p$-$n_e$ ($n_p$, $n_e$ have been estimated from experimentally found $R_H$ at 300 K).  The linear relation between $\delta v_\mathrm{FS}$ and $\delta n$ shown in Fig. 4(d) clearly testifies for the validity of our  approach to capture the experimental observations.

\section{Conclusions}
To summarize, we have synthesized and measured Hall effect on a series of ultra-thin NdNiO$_3$ films. The selective suppression of the simultaneous transitions in NdNiO$_3$ films through epitaxial strain engineering enable us to probe the paramagnetic metallic, paramagnetic insulating and antiferromagnetic insulating phases separately, without any detrimental influence of structural and charge ordering transitions.  This   approach reveals an unusual sign change in Hall coefficient across the $E^\prime$-type antiferromagnetic transition.  The appearance of such spin density wave transition from a paramagnetic insulating phase signals that the bond-disproportionation transition  creates a partially gapped Fermi surface.  Epitaxial strain also drastically changes the relative volume between hole and electron parts of the Fermi surface, resulting in a strain driven sign change in $R_H$  in the metallic phase of NNO.  Under compressive strain, all  of the  NNO films exhibit a non-Fermi liquid (NFL) behavior with  algebraic power exponents whereas  the film under  large tensile strain shows  magneto-transport behavior akin to the `strange metal' phase of  optimally doped high $T_c$ cuprates.  While superconductivity remains elusive  in the nickelate heterostructures  so far, these systems host several remarkable high $T_c$ cuprate signatures including   Zhang-Rice state, pseudogap, self-doping, NFL behavior with linear $T$ resistivity,  and spin-density wave.

\section{Acknowledgement}

SM acknowledges IISc start up grant, DST Nanomission (grant No. DST/NM/NS/2018/246), and SERB Early Career Research Award (ECR/2018/001512) for the financial support.  J.C. is funded  by Gordon and Betty Moore Foundation EPiQS Initiative through Grant GBMF4534. This research used resources of the Center for Functional Nanomaterials, which is a U.S. DOE Office of Science Facility, at Brookhaven National Laboratory under Contract No. DE-SC0012704

%\bibliographystyle{aipnum4-1.bst}

%\bibliographystyle{aipnum4-1}
%\bibliography{nickelatesref}
%

\end{document}